\documentclass[letter]{aa}  
\usepackage{graphicx}
\usepackage{txfonts}
\usepackage{color,soul}
\usepackage{tabularx}
\def\bd{J0331-27}
\def\xmm{{\em XMM-Newton}}

\begin{document}

   \title{EXTraS discovery of an X-ray superflare from an L dwarf}


   \author{Andrea De Luca
          \inst{1,2}
         \and
         Beate Stelzer \inst{3,4}
         \and
          Adam J.\ Burgasser \inst{5}
         \and
         Daniele Pizzocaro \inst{1}
         \and
          Piero Ranalli \inst{6}
         \and
         Stefanie Raetz \inst{3}
         \and
         Martino Marelli \inst{1}
         \and
         Giovanni Novara \inst{7}
         \and
         Cristian Vignali \inst{8,9}
         \and
         Andrea Belfiore \inst{1}
         \and
         Paolo Esposito \inst{7}
         \and
         Paolo Franzetti \inst{1}
         \and
         Marco Fumana \inst{1}
         \and 
         Roberto Gilli \inst{9}
         \and
         Ruben Salvaterra \inst{1}
         \and 
         Andrea Tiengo \inst{7,2}
          }

   \institute{INAF - Istituto di Astrofisica Spaziale e Fisica Cosmica Milano, Via A. Corti 12, 20133 Milano, Italy
             \and
             INFN, Sezione di Pavia, via A. Bassi 6, I-27100 Pavia, Italy
        \and
        Institut f\"ur Astronomie and Astrophysik T\"ubingen, Eberhard-Karls Universit\"at T\"ubingen, Sand 1, 72076, Germany
        \and INAF - Osservatorio Astronomico di Palermo, Piazza del Parlamento 1, 90134 Palermo, Italy
         \and
         Center for Astrophysics and Space Science, University of California San Diego, 9500 Gilman Drive, La Jolla, CA 92092, USA
         \and
         Combient mix AB, Kyrkogatan 22, 41115 Gothenburg, Sweden 
         \and
         Scuola Universitaria Superiore IUSS Pavia, Piazza della Vittoria 15, 27100 Pavia, Italy
         \and
         Dipartimento di Fisica e Astronomia, Universit\`a degli Studi di Bologna, Via P. Gobetti 93/2, 40129 Bologna, Italy
         \and
         INAF – Osservatorio di Astrofisica e Scienza dello Spazio di Bologna, Via P. Gobetti 93/3, 40129 Bologna, Italy
             }

\offprints{andrea.deluca@inaf.it}

   \date{Received -- ; accepted --}

 
  \abstract{We present the first detection of an X-ray flare from an ultracool dwarf of spectral class L. The event was identified in the EXTraS database of \xmm{} variable sources, and its optical counterpart, J0331-27, was found through a cross-match with the Dark Energy Survey Year 3 release. Next to an earlier four-photon detection of Kelu-1, J0331-27 is only the second L dwarf detected in X-rays, and much more distant than other ultracool dwarfs with X-ray detections (photometric distance of $240$\,pc). From an optical spectrum with the  VIMOS instrument at the VLT, we determine the spectral type of J0331-27 to be L1.  
  The X-ray flare has an energy of $E_{\rm X,F} \sim 2 \times 10^{33}$\,erg, placing it in the regime of superflares. No quiescent emission is detected, and from $2.5$\,Msec of \xmm{} data we derive an upper limit of $L_{\rm x,qui} < 10^{27}$\,erg/s. The flare peak luminosity ($L_{\rm x,peak} = 6.3 \times 10^{29}$\,erg/s), flare duration ($\tau_{\rm decay} \approx 2400$\,s), and plasma temperature ($\approx 16$\,MK) are similar to values observed in X-ray flares of M dwarfs. This shows that strong magnetic reconnection events and the ensuing plasma heating are still present even in objects with photospheres as cool as $\sim 2100$\,K.
  However, the absence of any other flares above the detection threshold of $E_{\rm X,F} \sim 2.5 \times 10^{32}$\,erg in a total of $\sim \, 2.5$\,Ms of X-ray data yields a flare energy number distribution inconsistent with the canonical power law $dN/dE \sim E^{-2}$, suggesting that magnetic energy release in J0331-27 -- and possibly in all L dwarfs -- takes place predominantly in the form of giant flares.}


   \keywords{X-rays: stars -- Stars: late-type -- Stars: activity -- Stars: flare -- Stars: coronae}

   \maketitle
   
   \titlerunning{X-ray Flaring L1 dwarf}
   \authorrunning{A. De Luca et al.}
%

\section{Introduction} \label{sec:intro}
Time-domain astronomy has uncovered 
a new class of stellar `superflares' with bolometric energies higher than $10^{33}$ erg \citep{maehara2012}, events that significantly influence planetary habitability \citep{armstrong2016,lingam2017} and may have been imprinted on meteoritic chemical abundances in our own solar system \citep{mishra2019}.

The most surprising superflare discoveries come from the  low-mass end of the  main sequence, the  ultracool dwarfs (UCDs) of spectral type L. Four white-light superflares from L dwarfs have been found in All Sky Automated Survey for Supernovae \citep{schmidt2016}, {\em Kepler} \citep{gizis2017,paudel2018} and Next Generation Transient Survey  data 
\citep{jackman2019}, each 
extremely bright (up to $\Delta V=-11$), releasing up to $4\times10^{34}$ erg of energy.
These bursts are comparable to the strongest flares in FGKM stars, and are 
remarkable given that most tracers of quiescent magnetic emission (H$\alpha$ and X-ray) decline in the L dwarf sequence \citep{schmidt2015,stelzer2006}. This decline is the result of increasingly neutral photospheres \citep{mohanty2002} and possibly of changes in magnetic reconnection \citep{mullan2010} or dynamo processes \citep{cook2014}. Furthermore, while young stars are typically the most magnetically active, three of the superflare L dwarfs appear to be older field stars.
What enables these dim stars to undergo one of the most dramatic stellar outbursts, and the frequency of such events, remains a mystery.

All of the L dwarf superflares reported to date have been detected in white-light observations.
In the standard flare scenario \citep[e.g.][]{cargill83} 
white-light flares are produced in the lower atmosphere following the bombardment with particles that are accelerated in magnetic reconnection events. The ensuing 
heating of the chromosphere leads to upwards plasma motions and, subsequently, to an X-ray flare from the hot matter confined in coronal magnetic loops.
Even for the Sun, observationally associating X-ray and optical flares has remained difficult \citep{hao17}.
For L dwarfs the lack of simultaneous multi-band observations has so far prevented tests of the solar flare scenario.


Here we report on the first detection of an X-ray superflare emitted by a distant ($\sim240$ pc) 
early L-type dwarf. 
Previously, only one other L dwarf had been detected in (quiescent) X-rays, the binary system Kelu 1AB at 19 pc \citep{audard2007}.
Our results were obtained within the context of    {\em Exploring the X-ray Transient and variable Sky\footnote{\url{www.extras-fp7.eu}}} (EXTraS), an EU-funded project aimed at extracting and characterising  all temporal information stored in serendipitous {\em XMM-Newton} data \citep[][De Luca et al., in preparation]{deluca17}.  


This letter is organised as follows: in Sect.~\ref{sec:selection} we describe the discovery of the flare; Sects.~\ref{sec:flare} -- \ref{sec:star} give temporal and spectral properties of the flare, set constraints on the flaring rate and on the quiescent X-ray emission of the source, and give results on the spectral classification of the dwarf and on its properties. 
All results are discussed in Sect.~\ref{sec:disc}.

\section{The X-ray flare of J0331-27} \label{sec:selection}
A large catalogue listing 11945 candidate ultracool dwarfs of the L and T spectral classes was recently published by \citet{carnerorosell2019}. This catalogue is based on photometric classification of data from the Dark Energy Survey Year 3 release \citep{des2016} covering $\sim2400$ square degrees down to $i_{AB}=22$, matched to photometry from the Vista Hemisphere Survey \citep{mcmahon2013} and Wide-field Infrared Survey Explorer \citep{wright2010,mainzer2011}. Of the sources of this catalogue, 515 are located within the field of view of at least one {\em XMM-Newton} observation. Cross-correlation of the catalogue 
with the EXTraS database singles out 3XMM\,J033158.9-273925 (hereafter J0331-27) as a very interesting case. J0331-27 matches within 1$\farcs$1 the position of the L0 candidate Obj.\
ID\ 366318917   ($\rm   RA=03^h31^m59\fs07$, $\rm Dec=-27^{\circ}39'25\farcs7$; J2000) in \citet{carnerorosell2019}, and is listed as a variable X-ray source in the EXTraS database. Inspection of light curves in the EXTraS Public Archive\footnote{\url{https://www88.lamp.le.ac.uk/extras/archive}} clearly shows an X-ray flare.
Moreover, the source is located within the Extended {\em Chandra} Deep Field South \citep[ECDFS,][]{lehmer2005}, one of the most scrutinised portions of the sky, with a large amount of deep multi-wavelength data available. 



\section{Temporal and spectral properties of the flare} \label{sec:flare}
\begin{figure}
   \centering
   \includegraphics[width=6cm,angle=-90]{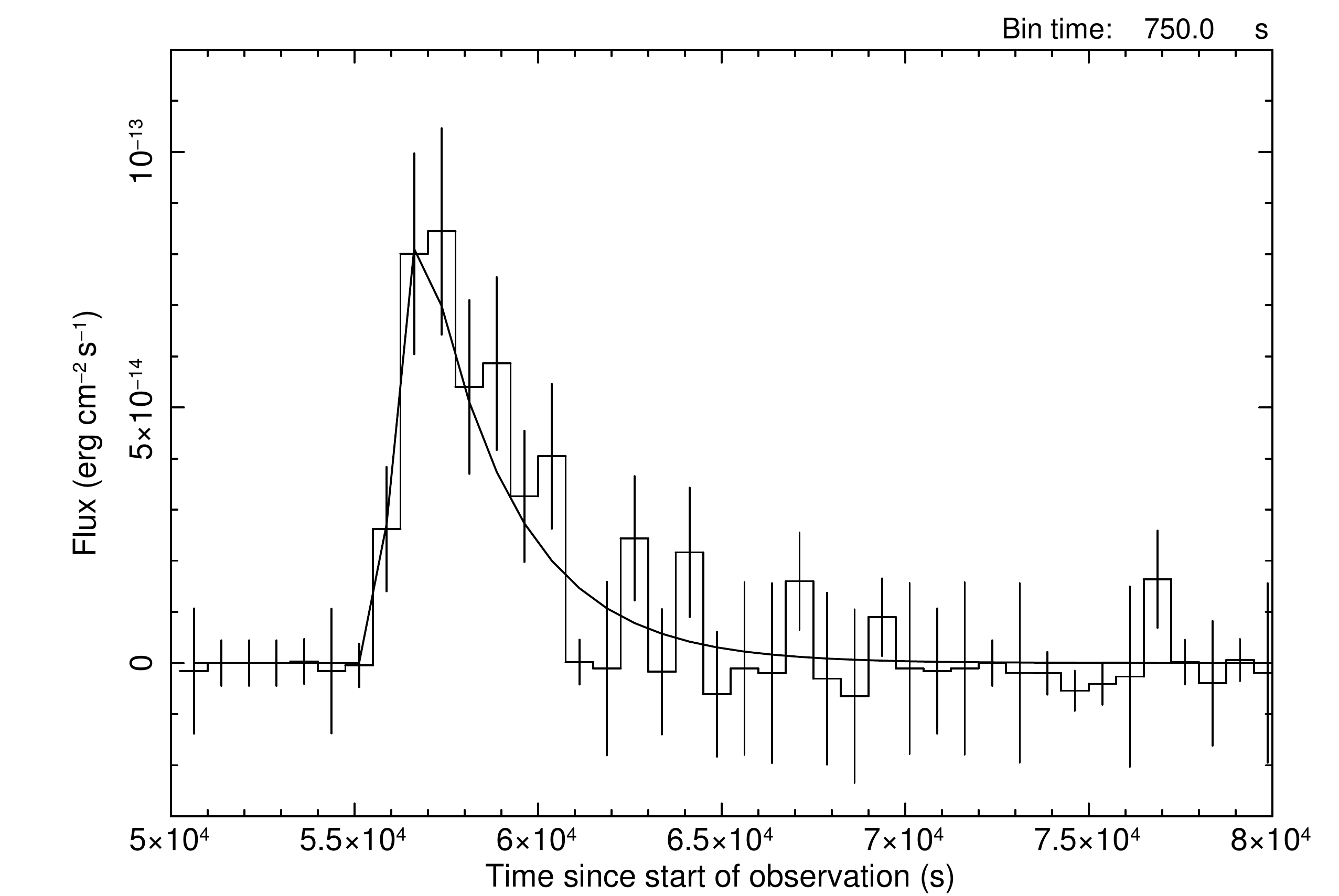}
      \caption{Background-subtracted light curve of J0331-27 in the 0.2-2 keV energy range combining pn and MOS data. A 30 ks portion of the observation is shown. Count rate has been converted to flux in the 0.5-2 keV energy range using the best fit spectral model (see text). The flare from the source is apparent, and is nicely described by a linear rise and exponential decay model (see text).}
         \label{epiclc}
   \end{figure}

The X-ray flare from J0331-27 is clearly seen in data collected on 2008 July 5 by all detectors of the European Photon Imaging Camera (EPIC) instrument on board {\em XMM-Newton}, namely the pn camera \citep{Strueder2001} and the two MOS cameras \citep{Turner2001}. The observation (Obs. ID 0555780101) lasted 130 ks and was performed as a part of the {\em XMM-Newton} Ultra-deep Survey of the {\em Chandra} Deep Field South \citep[XMM-CDFS;][]{comastri11,ranalli2013}. 
In Fig.~\ref{epiclc} we show a background-subtracted light curve combining pn and MOS data, generated using an updated version of the EXTraS software \citep{marelli17}.  The flare profile is nicely described (reduced $\chi^2=0.88$, 35 d.o.f.) by a linear rise (rise time of $1600\pm300$ s), peaking at MJD=$54652.8939\pm0.0032$ UTC (in Barycentric Dynamical Time), followed by an exponential decay of the form $e^{-t/\tau_{decay}}$ with $\tau_{decay}=2400\pm400$ s. 
The peak flux is $(9.0\pm1.5)\times10^{-14}$ erg cm$^{-2}$ s$^{-1}$ and the integrated flare energy flux ({\em fluence}), evaluated by integrating the model of the flare time profile, is $\sim2.7\times10^{-10}$ erg cm$^{-2}$ (all values refer to the 0.5-2 keV energy range).
The best fit model has a continuum level consistent with zero;  no significant emission is seen apart from the flare. 

To perform spectroscopy of the flare, we extracted source counts from a circle with radius of $15^{\prime\prime}$ and background counts from a nearby, source-free region. We only considered observing times starting 3000 s before the peak of the flare and extending up to 12000 s after the peak. Time intervals with high backgrounds were  excluded following the prescription by \citet{delucamolendi2004}. 
The resulting flare spectrum contains $100\pm12$ background-subtracted counts from the three EPIC detectors. We generated a response matrix and effective area file using the SAS tasks \texttt{rmfgen} and \texttt{arfgen}.  Spectral modelling was performed using the XSPEC v12.10 Software. A good description of the data is given by an optically thin thermal plasma model (apec) with kT$=1.39^{+0.27}_{-0.11}$ keV and fixed abundance assumed to be 0.3 solar. The absorbing column is consistent with zero (N$_H<1.7\times10^{20}$ cm$^{-2}$ at 1$\sigma$); total Galactic absorption in the direction of the target is indeed very low \citep[N$_{H,Gal}=6\times10^{19}$ cm$^{-2}$,][]{benbekhti2016}. The goodness of the fit, evaluated as the percentage of Monte Carlo realisations that had Kendall's W-statistic values lower than that of the best fit, is 84\% (based on $10^4$ simulations).

The field of J0331-27 was observed 33 times by {\em XMM-Newton}. 
We took advantage of this large dataset (total nominal exposure of $3.45$\,Ms) to search for other flares from J0331-27. 
No flares were detected within the blind search for transient sources performed by EXTraS, 
with duration $<10000$ s.
To set an upper limit to the integrated energy flux of unseen flares, 
we evaluated the average background level close to the position of the source in individual observations. 
We followed \citet{ranalli2013} in their steps for data reduction, event filtering (including soft proton flare screening), and astrometry correction. We excluded observation 0555780101, in which the flare was detected. Images and exposure maps were generated for each camera and for each observation in the 0.5-2 keV energy range. As a sanity check, we performed aperture photometry by extracting source counts from a circle of $10^{\prime\prime}$ radius and background counts from a contiguous circle with the same radius, combining results from all EPIC cameras for each observation;  in no case was significant emission detected from J0331-27. Based on the observed average background rate (assuming a Poissonian regime) and adopting the spectral model best fitting the observed flare of J0331-27, we estimated that an integrated energy flux of $\sim4\times10^{-11}$ erg cm$^{-2}$ in 5 ks would have produced a $3\sigma$ excess of counts, which would have triggered the detection algorithm in EXTraS.   
We note that the 3XMM catalogue lists a detection of J0331-27 in dataset 0108060701. However, the source is detected at the $3\sigma$ level only in the MOS1 hard band  (4.5-12 keV), with upper limits at softer energies not consistent with a thermal spectrum. 
We disregard this detection as an artefact.

\section{Search for X-ray quiescent emission} \label{sec:qemiss}


We  searched  for  quiescent  emission  from  J0331-27  by  stacking  all  data  from  the  XMM-CDFS  survey,  excluding   Obs. ID 0555780101 ($\sim2.5$ Ms good observing time after screening for soft proton flares). Considering only PN data and the 0.5–2 keV band, we extracted 362 counts for J0331-27 from a circular region with a 10$''$ radius and 349 background counts from an adjacent circular region with the same radius. Assuming a Poisson background distribution, the observed counts can be safely attributed to a background fluctuation.

An upper limit to the quiescent flux of J0331-27 can be calculated by requiring that a detection with the PN camera had at least 50 more counts than in the background region. Formally, such an excess of counts would only have a 0.4\% probability of being due to  a background fluctuation. In practice, requiring an excess of 50 counts over the background is consistent with what is observed for other off-axis sources in the XMM-CDFS catalogue \citep{ranalli2013}.  
To convert counts to flux, we generated response matrices and effective area files for all XMM-CDFS observations (again excluding Obs. ID 0555780101) and averaged them following \citet{georgantopoulos2013}. Assuming the same  spectral model as observed during the flare, we calculate the upper limit to the quiescent emission as  $1\times 10^{-16}$ erg~s$^{-1}$~cm$^{-2}$ in the 0.5-2 keV energy range.


The field of J0331-27 was observed several times by {\em Chandra}. 
It is included in the Extended CDFS, for a total of $\sim250$\,ks (Obs. ID 5017 and 5018, PI Brandt). The source was not detected by \cite{lehmer2005}. We retrieved {\em Chandra} data  and analysed them with the CIAO v4.11 software and CALDB v4.8.3. We reprocessed level 2 data using the \texttt{chandra\_repro} script and merged the resulting event files using the  \texttt{reproject\_obs} script. Using the \texttt{srcflux} script and adopting the spectral shape of the flare emission, we  found a $3 \sigma $ upper limit of $5\times10^{-16}$ erg cm$^{-2}$ s$^{-1}$ to the flux of J0331-27 in the 0.5-2 keV energy range.
The position of J0331-27 lies also very close to the edge of the field of view in the CDFS 7Ms survey, at an off-axis angle of $\sim11$ arcmin, covered with an exposure time (corrected for vignetting) of $\sim400$ ks. The source is not detected in these data \citep{luo2017}. We retrieved the image and exposure map of the survey in the 0.5-2 keV energy range made available by the CDFS team\footnote{\url{http://personal.psu.edu/wnb3/cdfs/cdfs-chandra.html}}. Following \citet{lima1983}, we computed the count rate needed to produce a $3\sigma$ excess in the image at the position of J0331-27 within the PSF area and converted it into flux assuming the observed spectral shape of the flare. The resulting upper limit to the source flux is also $\sim5\times10^{-16}$ erg cm$^{-2}$ s$^{-1}$.

\section{Classification and properties of the UCD} \label{sec:star}

\vspace{0.5cm}
\subsection{Spectral classification}
\begin{figure}
   \centering
   \includegraphics[width=9cm]{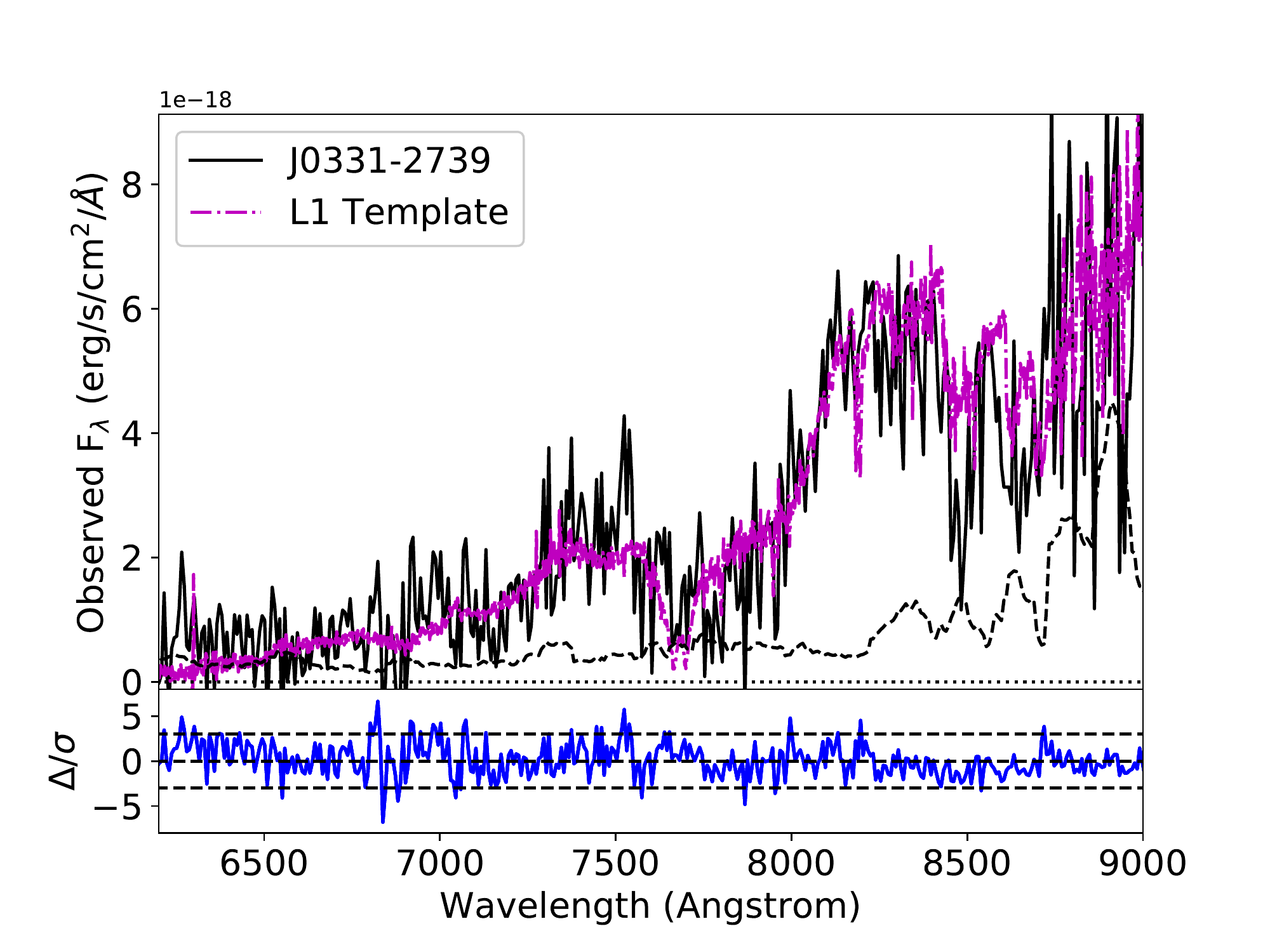}
      \caption{{\em Top panel}: VIMOS spectrum of J0331-27 (black line) compared to an L1 SDSS spectral template from 
      \citet[magenta dot-dashed line]{2014PASP..126..642S}. The spectral data uncertainty is shown as the dashed line.
      {\em Bottom panel}: Difference between the observed data and template ($\Delta$) normalised to the uncertainty ($\sigma$). }
         \label{spectrum}
   \end{figure}

An optical spectrum of J0331-27 was observed with the Visible Multi Object Spectrograph (VIMOS) instrument \citep{2003SPIE.4841.1670L} as part of the VIMOS/VLT Deep Survey of the CDFS \citep{2005A&A...439..845L}, where the source is listed with Object ID 38073.
Data were reduced using the automatic pipeline developed by the survey team. The spectrum (Figure~\ref{spectrum}) shows the characteristic steep rise in continuum across the optical band, with strong absorption features of K~I, TiO, VO, FeH, CrH apparent, typical of late-M and L dwarfs \citep{1999ApJ...519..802K}. This spectrum was compared to red optical templates of M and L dwarfs compiled from Sloan Digital Sky Survey (SDSS) optical spectroscopy by \citet{2007AJ....133..531B} and \citet{2014PASP..126..642S} in the 6500-8000~{\AA} range using a $\chi^2$ statistic. The best fit templates were L1 and (as a close second) L0, with a weighted mean of L0.7$\pm$1.3, consistent with the photometric classification of \citet{carnerorosell2019}\footnote{ The source had been previously classified as an M6 dwarf, or as an unresolved elliptical galaxy at $z\sim1.9$, by \citet{groenewegen2002} based on five-band (UBVRI) photometry and morphology (source ID J033159.06-273925.5 in their catalogues). It was also classified as a galaxy by \citet{wolf2008}, based on multi-band photometry (Obj. 51627 in their catalogue).}. There is no evidence of H$\alpha$ emission, and low signal-to-noise ratio of the data made it impossible to determine the presence of Li~I absorption (a diagnostics of substellarity) or features indicative of low surface gravity \citep{cruz2009}.

\vspace{0.5cm}
\subsection{Photometric distance}\label{sect:dist}
 \bd{} is not listed in Gaia Data Release 2, thus we have to rely on photometric distance estimates.
\cite{carnerorosell2019} 
 found $\sim280$ pc 
from the comparison 
between the observed near-infrared (NIR) magnitudes and the absolute magnitudes for L0 subtype.
Absolute magnitudes are anchored to $M_{\rm W1}$ and $M_{\rm W2}$ from Table 14 of \cite{dupuyliu2012}. The other absolute magnitudes are obtained from the colours in Table 3 of \cite{carnerorosell2019}. Applying the same procedure under the   hypothesis of spectral type L1 and averaging over all available bands we obtain  a distance of 
$240_{-20}^{+40}\,{\rm pc}$. 
This is the value we adopt throughout this paper.

\subsection{Age characterisation}
The age of J0331-27 is of interest as younger stars tend to be more active, whereas the white light superflare UCDs identified to date appear to be older field objects \citep{schmidt2016}. Age determinations for individual UCDs are challenging, however, as traditional age metrics such as 
(quiescent) magnetic activity level and rotation rates appear to decouple from age at the lowest stellar masses and coolest temperatures (e.g. \citealt{2011ApJ...727...56I,schmidt2015}). Surface gravity-sensitive features in the optical spectra of L dwarfs, such as enhanced VO band and weakened alkali line absorption \citep{2008ApJ...689.1295K,cruz2009} provide approximate age constraints for sources younger than $\sim$300~Myr; unfortunately, the VIMOS spectrum of J0331-27 is too noisy to discern these features. Similarly, kinematics cannot be used as the VIMOS data cannot provide radial velocity information, and no proper motion has been reported for this source. 

Colour provides an additional age diagnostic. 
\cite{faherty2016} show that NIR colours can be used to distinguish between low-gravity and field-gravity dwarfs, with the latter appearing systematically redder than the mean of the field dwarfs at a given spectral type. Table~\ref{tab:colors} compares the colours of J0331-27 to the average colours of low surface gravity and field L1 dwarfs reported in \cite{faherty2016}.  Rather than being systematically redder, J0331-27 is   bluer than field L1 dwarfs, particularly in $J-K_s$ and $J-W1$ colours, consistent with older (high surface gravity) and slightly metal-poor L dwarfs \citep{2018MNRAS.480.5447Z}.  
Thus, it is most likely that J0331-27 is an older field L dwarf. 

\begin{table}[!tb]
\centering
\caption{Infrared colours of J0331-27 compared to mean colours of the low-gravity and field-gravity L1 dwarfs from  \citep{faherty2016}.}
\label{tab:colors}
\begin{tabular}{lccc} \hline
Colour & \bd & <Low-g> & <Field>\\ \hline
${\rm J-H}$ &$0.61\pm0.15$ &$0.94\pm 0.06$& $0.81\pm0.14$\\
${\rm J-K_s}$ &$1.10\pm0.18$ &$1.61\pm0.13$ & $1.35\pm0.19$\\
${\rm J-W1}$ & $1.27\pm0.19$&$2.17\pm0.20$ & $1.71\pm0.21$\\
${\rm J-W2}$ &$1.80\pm0.33$ & $2.55\pm0.25$& $1.97\pm0.23$\\
${\rm H-K_s}$ &$0.49\pm0.19$ &$0.67\pm0.11$ &$0.54\pm0.13$\\
${\rm H-W1}$ &$0.66\pm0.20$ &$1.21\pm0.17$ & $0.91\pm0.15$\\
${\rm H-W2}$ &$1.20\pm0.34$ &$1.60\pm0.22$ & $1.17\pm0.18$\\
${\rm K_s-W1}$ &$0.16\pm0.23$ &$0.54\pm0.11$ &$0.37\pm0.11$\\
${\rm K_s-W2}$ &$0.70\pm0.37$ &$0.92\pm0.14$ & $0.63\pm0.14$\\
${\rm W1-W2 }$&$0.54\pm0.38$ &$0.38\pm0.05$ & $0.26\pm0.06$\\ \hline
\end{tabular}
\end{table}

\begin{figure}[!htb] 
\centering
\includegraphics[width=9.1cm]{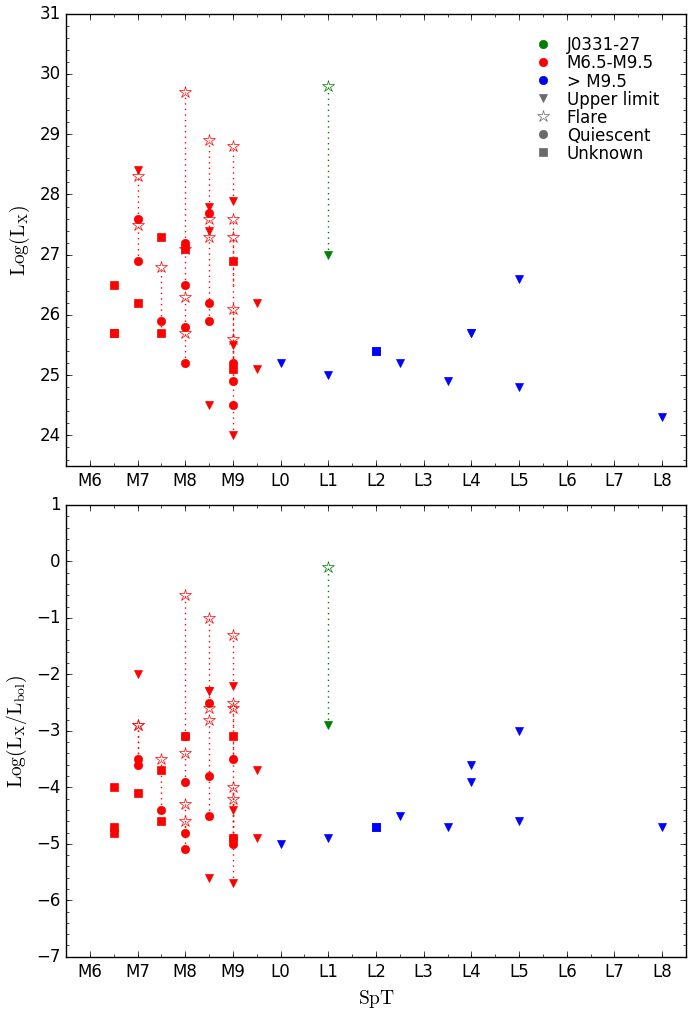}
\caption{$L_X$ and $L_X/L_{bol}$ vs. spectral type for \bd{} (flare peak luminosity and upper limit on the quiescent luminosity), compared with the other UCDs for which X-ray data are available (from \citealt{stelzer2006}, \citealt{williams2014}, \citealt{cook2014} and references therein; \citealt{robrade2010} and \citealt{gupta2011}).
} 
\label{fig:ucds_xray}
\end{figure}

\section{Discussion} \label{sec:disc}

%

The association of the flaring source J0331-27 with the L1 dwarf is robust. 
First, the chance alignment probability is 
low ($<2\times10^{-4}$). Second,
archival {\em HST} images show that the dwarf is the only point-like source within the error circle of the X-ray source.
Assuming the two sources to be unrelated would require that the flaring source counterpart have m$_V>27$ and m$_{z}>26$. This requires a Galactic flaring X-ray source {extremely dim} at optical--NIR wavelengths, or an extragalactic transient. Both cases would require peculiar and unlikely scenarios to explain the X-ray emission and temporal behaviour.

 The available data do not allow us to precisely constrain  the age of \bd{}, although its IR colours indicate that  it is likely an older source. Assuming an age $\gtrsim1$\,Gyr, evolutionary models \citep{2003A&A...402..701B,2001RvMP...73..719B} predict the mass of \bd{} to be above the hydrogen burning minimum mass limit (M $>$ 0.072~M$_{\odot}$).

In Fig. \ref{fig:ucds_xray} we put the X-ray emission of J0331-27 in context to that of other UCDs.
J0331-27 is the first L dwarf to be detected in an X-ray flare. We measure 
a flare peak luminosity of 
$\log{L_{\rm x,peak}}=29.8\,{\rm (erg/s)}$  and  $\log{(L_{\rm x,peak}/L_{\rm bol})}=-0.1$,
similar to X-ray flares observed on late-M dwarfs (e.g. \citealt{stelzer2006b}, \citealt{hambaryan2004}, \citealt{gupta2011}) and actually somewhat larger than the majority of them. 
The quiescent luminosity is only weakly constrained with an upper limit of  
$\log{L_{\rm x,qui}}<27.0\,{\rm (erg/s)}$
 and ${\log{(L_{\rm x,qui}/L_{\rm bol})}<-2.9}$, a result of the large distance of \bd{} compared with the other UCDs studied in X-rays so far. 
Similarly, an upper limit on the radio flux of J0331-27 of $ 7.4\,\mu$Jy at 1.4GHz \citep{miller2013}, translates into a very weak constraint on the radio luminosity, $ \log{\rm   L_{\nu,R}}<14.7$ (erg s$ ^{-1}$ Hz$ ^{-1}$),  preventing analysis of the well-established deviation of UCDs from the G\"udel-Benz radio/X-ray relation \citep{guedelbenz93,berger2002}.

The short observed flare duration (decay timescale $\tau \simeq 2400\,{\rm s}$) is also consistent with those of the 
 X-ray flares on late-M dwarfs. This indicates a compact size for the flaring region on the UCD surface (see \citealt{stelzer2006b}).
Finally, 
the observed X-ray emitting plasma temperature of $\sim 16\,{\rm MK}$
is  within the range of values 
reported for late-M dwarf flares. In summary, our observation shows that no qualitative change takes place in the properties of X-ray flares at the bottom of the main sequence down to 
$T_{\rm eff} \sim 2100$\,K. 



 The remarkable presence of a single superflare in 
$\sim2.5$\,Ms of {\em XMM-Newton} data 
 gives rise to the question on the frequency of such events and of X-ray flares in general on L-type dwarfs. The integrated X-ray flare energy of the event on \bd{}  is  
$\log{E_{\rm X,F}}=33.3\,{\rm(erg)}$, 
and the flare frequency is $\nu (\log{E_{\rm X,F}} \gtrsim 33.3) \sim 1 / 30\,{\rm d^{-1}}$. 
Based on the sensitivity of the available {\em XMM-Newton} observations (assuming a typical flare timescale of 5,000 s, see Sect.~\ref{sec:flare}) we estimate that $\sim$eight flares above $\log{E_{\rm X,F}} \sim 32.4\,{\rm (erg)}$ would have been detected if \bd{} obeyed the canonical power law for the flare energy number distribution, $\frac{dN}{dE_{\rm F}} \sim E_{\rm F}^{\alpha}$ with $\alpha \approx -2$ \citep[see references in][]{Argiroffi19.0}. The fact that these flares are not seen suggests a non-standard flare energy distribution for \bd{}.

Contrary to the large flare on the M8 dwarf LP412-32, which was observed simultaneously with the \xmm{} X-ray instruments and its Optical Monitor \citep{stelzer2006b}, no contemporaneous optical data is available for the event on \bd{}. However, in view of the similarities of the \bd{} X-ray flare and the X-ray flares on late-M dwarfs described above, we can use the observed optical-to-X-ray energy ratio of LP412-32    
($E_{\rm opt,F} \approx E_{\rm X,F}$)
to estimate an optical counterpart of $E_{\rm opt,F} \gtrsim 10^{33}$\,erg for the X-ray superflare on \bd{}. 
Remarkably, other white-light superflares observed on L dwarfs (without  simultaneous X-ray data)  show flare energies of the same order (\citealt{jackman2019}, \citealt{gizis2017}). 
We note that for a flare on an early M dwarf observed simultaneously with {\em XMM-Newton} and {\em Kepler} \citep[KIC\,8454353; see][]{Pizzocaro19.0} we also find that the X-ray and optical flare energy are within a factor of two of each other. On the other hand,  \cite{Guarcello19.0} find significantly higher emission in the optical {\em Kepler} band with respect to X-rays for some flares in the Pleiades.
Simultaneous optical--X-ray studies of 
larger samples are required to nail down the relative radiative output of chromosphere and corona, and whether this depends on other stellar parameters.



Systematic searches are also likely to yield better constraints on the frequency of X-ray flares on L dwarfs. Our
cross-correlation of the catalogue of UCD candidates by \citet{carnerorosell2019} with the \xmm{} serendipitous source catalogue \citep[3XMM-DR8\footnote{\url{http://xmmssc.irap.omp.eu/Catalogue/3XMM-DR8/3XMM_DR8.html}},][]{rosen16} yielded two additional close matches between an L dwarf candidate and an X-ray source\footnote{3XMM J205205.8-610355, within $\sim1.6''$ of 
Obj. ID 185442097 in \citet{carnerorosell2019} and 3XMM J232604.0-543340, within $\sim1.0''$ 
Obj.\ ID 133082583 in \citet{carnerorosell2019}}. 
Interestingly, both X-ray sources display possible flaring activity, as seen in EPIC light curves produced by the EXTraS software. Optical--NIR spectroscopy 
is
needed to confirm the UCD classification
of these  objects.
Finally, we estimate that the upcoming All-Sky Survey of {\em eROSITA} \citep{Merloni12.0} will be able to detect superflares of the size of the event on \bd{} described here within a volume of $\sim 100$\,pc.

\begin{acknowledgements}
This work is based on observations obtained with XMM-Newton, an ESA science mission with instruments and contributions directly funded by ESA Member States and NASA. This research has made use of data produced by the EXTraS project, funded by the European Union's Seventh Framework Programme under grant agreement no 607452. We also used observations made by the Chandra X-ray Observatory, and obtained from the Chandra Data Archive. We thank Gian Luca Israel for useful discussions. We  acknowledge  the computing  centre  of  INAF Osservatorio  Astrofisico  di Catania for the availability of computing resources and support under the coordination of the CHIPP project. We acknowledge financial support from ASI under ASI/INAF agreement N.2017-14.H.0.
\end{acknowledgements}


\end{document}